\documentclass[english,twocolumn]{article}
\usepackage[utf8]{inputenc}
\usepackage[english]{babel}
\usepackage[big,online]{dgruyter}
\usepackage[sort&compress,square,numbers]{natbib}
\usepackage{bm}
\usepackage{amsmath,amsfonts,amssymb}
\usepackage{txfonts}
\usepackage{tabularx}
\usepackage{textcomp}
\usepackage{microtype}
\usepackage[separate-uncertainty=true]{siunitx} 

\begin{document}

 \journalname{Current~Directions~in~Biomedical~Engineering}
 \journalyear{2021}
 \journalvolume{tbd}
 \journalissue{0}
 \startpage{0}
 \DOI{...}
 \openaccess
 \contributioncopyright[]{2021}{}{}

  \author*[1]{R. Mieling}
  \author[1]{J. Sprenger}
  \author[1]{S. Latus}  
  \author[1]{L. Bargsten} 
  \author[1]{A. Schlaefer}
  
  \runningauthor{Mieling et al.}
  \affil[1]{\protect\raggedright Institute of Medical Technology and Intelligent Systems, Hamburg University of Technology, Hamburg, Germany, \newline e-mail: robin.mieling@tuhh.de}
  \title{A novel optical needle probe for deep learning-based tissue elasticity characterization}
  \runningtitle{A novel optical needle probe for deep learning-based tissue elasticity characterization}

\abstract{The distinction between malignant and benign tumors is essential to the treatment of cancer. The tissue's elasticity can be used as an indicator for the required tissue characterization. Optical coherence elastography (OCE) probes have been proposed for needle insertions but have so far lacked the necessary load sensing capabilities. 

We present a novel OCE needle probe that provides simultaneous optical coherence tomography (OCT) imaging and load sensing at the needle tip. We demonstrate the application of the needle probe in indentation experiments on gelatin phantoms with varying gelatin concentrations. We further implement two deep learning methods for the end-to-end sample characterization from the acquired OCT data. 

We report the estimation of gelatin sample concentrations in unseen samples with a mean error of $1.21 \pm 0.91$ wt\%. Both evaluated deep learning models successfully provide sample characterization with different advantages regarding the accuracy and inference time.}


\maketitle

\section{Introduction}
The response to mechanical stress significantly varies for different tissue types and cancerous tissues exhibit different elasticities compared to their healthy counterpart~\cite{good2014elasticity}. Tissue elasticity can therefore serve as a biomarker for tissue characterization. Clinicians can feel for stiff inclusions associated with pathology but manual palpation is subjective and feedback on local forces is difficult to acquire in modern minimally invasive surgery (MIS). Instead, image-based elastography has been proposed for the measurement of elasticity in biological tissue by mapping local deformations to an applied mechanical load. Ultrasound elastography~\cite{cui2015endoscopic} and magnetic resonance elastography~\cite{papazoglou2012multifrequency} have been implemented for tissue characterization at a resolution of hundreds of micrometers. 
In recent years, Optical coherence elastography (OCE) has gained attention as an extension to optical coherence tomography (OCT). OCT can help in the detection of micrometer structures due to its high spatial resolution of \SIrange[]{1}{10}{\micro\meter}~\cite{kennedy2013review}. OCE is commonly limited to the application to superficial target regions due to the maximum imaging depth of approximately \SI{2}{\milli\meter}. %
But OCE can be integrated in a needle probe to extend the application to deeper target regions. Needle-based OCE has been proposed with shear wave~\cite{latus2017approach} and compression loading~\cite{kennedy2013needle}. Compression-based OCE offers higher lateral resolution but the lack of local load sensing within the previously proposed needle-probe only provides qualitative elasticity measurements. Friction forces along the needle shaft are superimposed with the local tip forces and a dedicated force sensor located at the needle tip is required. Tissue characterization is only realizable with a known relation to the applied load. So far, sensors for the measurement of needle tip forces~\cite{beekmans2017fiber} as well as needle-based OCT probes for guided interventions~\cite{carrasco2017review} have been proposed exclusively. 

In this work, we present a novel compression-based OCE needle-probe with simultaneous load sensing and imaging capabilities. We further propose the direct tissue characterization for compression-based OCE via deep learning and demonstrate our methods on tissue mimicking gelatin gels. A similar approach has recently shown success in shear wave OCE~\cite{neidhardt20204d}. The end-to-end characterization removes the need for intermediate calculations of local displacement fields that are otherwise needed for compression-based OCE. 

\begin{figure*}[t]
    \centering
    \begin{minipage}{.40\textwidth}
        \centering
	    \includegraphics[width =\textwidth]{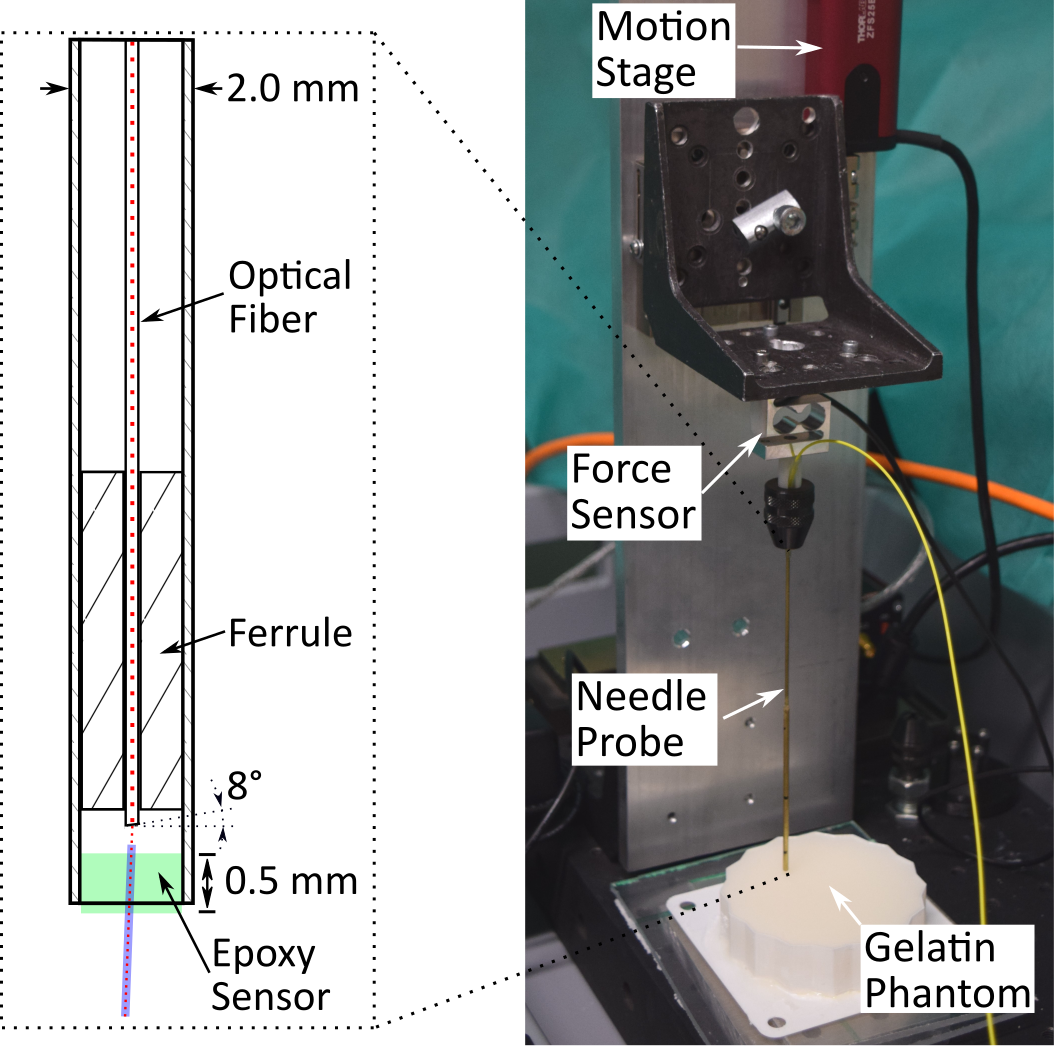}
	\end{minipage}%
	\hspace{0.1\textwidth}
	\begin{minipage}{.46\textwidth}
        \centering
	    \includegraphics[width =\textwidth]{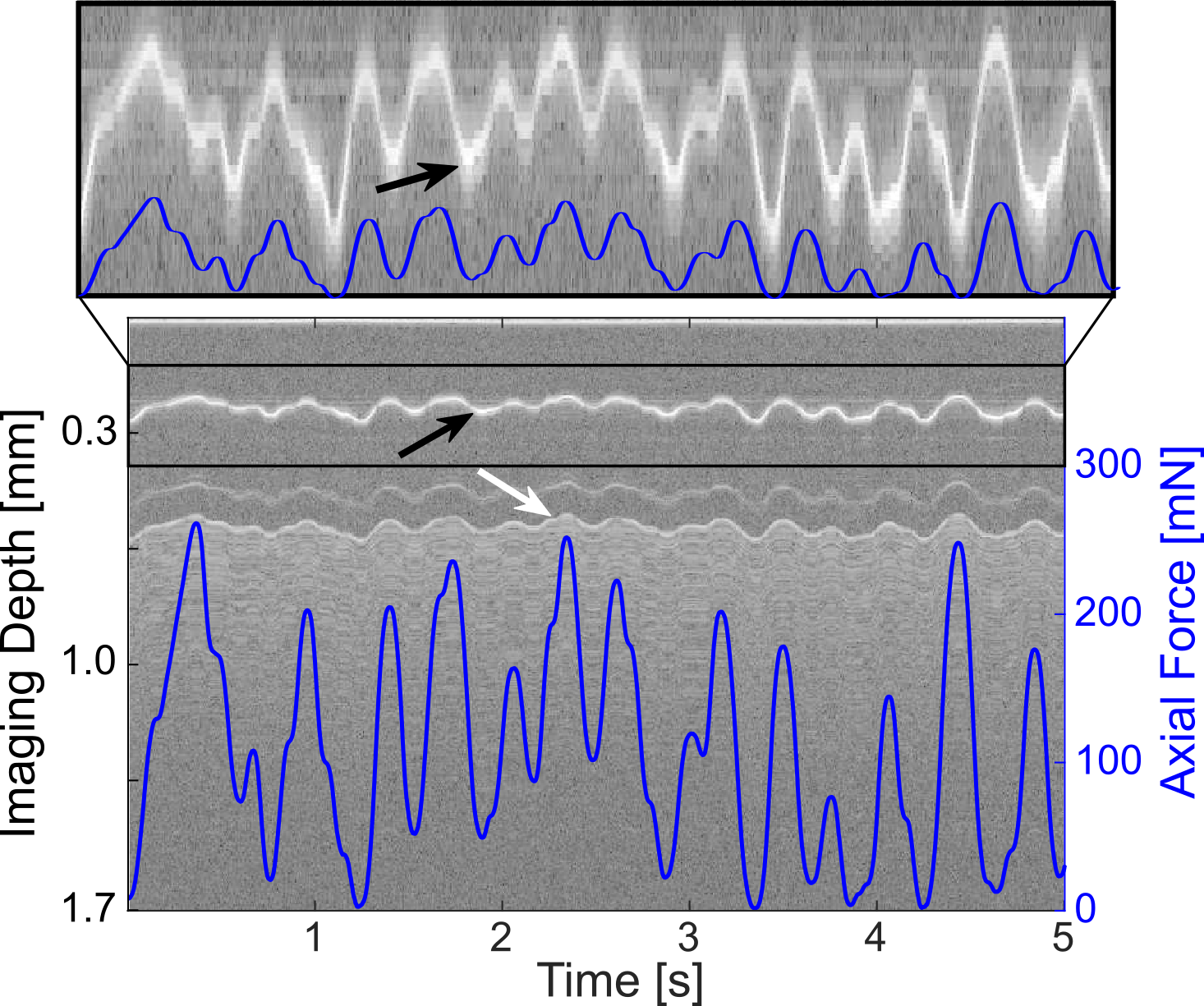}
	\end{minipage}%
	\caption{\textit{Left}: Experimental setup with a schematic of the proposed needle probe. A force sensor and motion stage were used for the indentation of the gelatin samples. The depth scan of the OCT system (blue line) visualized deformations of the epoxy sensor (green) and the sample beyond. \textit{Right}: Example recorded during the indentation of a single phantom. The deformation of the sensor (interfaces marked by black and white arrow) provides information on the local tip load while the sample is simultaneously imaged under deformation.}
	\label{fig:needle}
\end{figure*}


\section{Materials and Methods} 

\subsection{System Setup and OCT Data}
Our custom built needle probe was designed as illustrated in Figure~\ref{fig:needle}. A single-mode glass fiber (SMF-28,Thorlabs GmbH,GER) was embedded into a hollow needle and fixed via a ferrule. A flat tip needle was used for the compression of the sample and the distal end was fitted with a cylindrical sensor. The sensor was cast out of a translucent epoxy resin (NOA 63 and 1625, Norland Products Inc., USA). The end of the optical fiber was angled at \ang{8} to minimize common-path reflections and provided a forward facing view that visualized both the sensor and the sample. The needle probe was used for the indentation of gelatin gels with simultaneous OCT imaging under load. Axial scans (A-scans) were acquired at a sampling rate of~\SI{5.5}{\kilo\hertz} via the spectral domain OCT imaging system (Telesto Telesto I, Thorlabs GmbH, GER). The flexible sensor deformed upon indentation of the sample and the resulting movement is proportional to the locally occurring load. A high precision, linear translation stage (ZFS25B, Thorlabs GmbH, GER) and a uniaxial force sensor (KD24s, ME-Meßsysteme GmbH, GER) were used for the indentation of tissue mimicking gelatin gels. Different weight ratios [wt\%] between gelatin and water were used to produce the phantoms. The weight ratio was used as a surrogate label for the elasticity and the two terms are used interchangeably in the context of this work. 

\subsection{Deep Learning Problem}
We considered the end-to-end learning problem from sequentially acquired 1D A-scans to the direct sample characterization utilizing the varying resistance to deformation for each sample elasticity. An image $M_i \in \mathbb{R}^{nxm}$ assembled from a temporal sequence $n$ of consecutive A-scans $A_i \in \mathbb{R}^m$ was directly mapped to the gelatin weight ratio of the imaged sample under load. It was considered as a continuous variable and the sample characterization was consequently handled as a regression problem. 

\begin{figure*}[t]
	\centering
	\includegraphics[width = 0.85\textwidth]{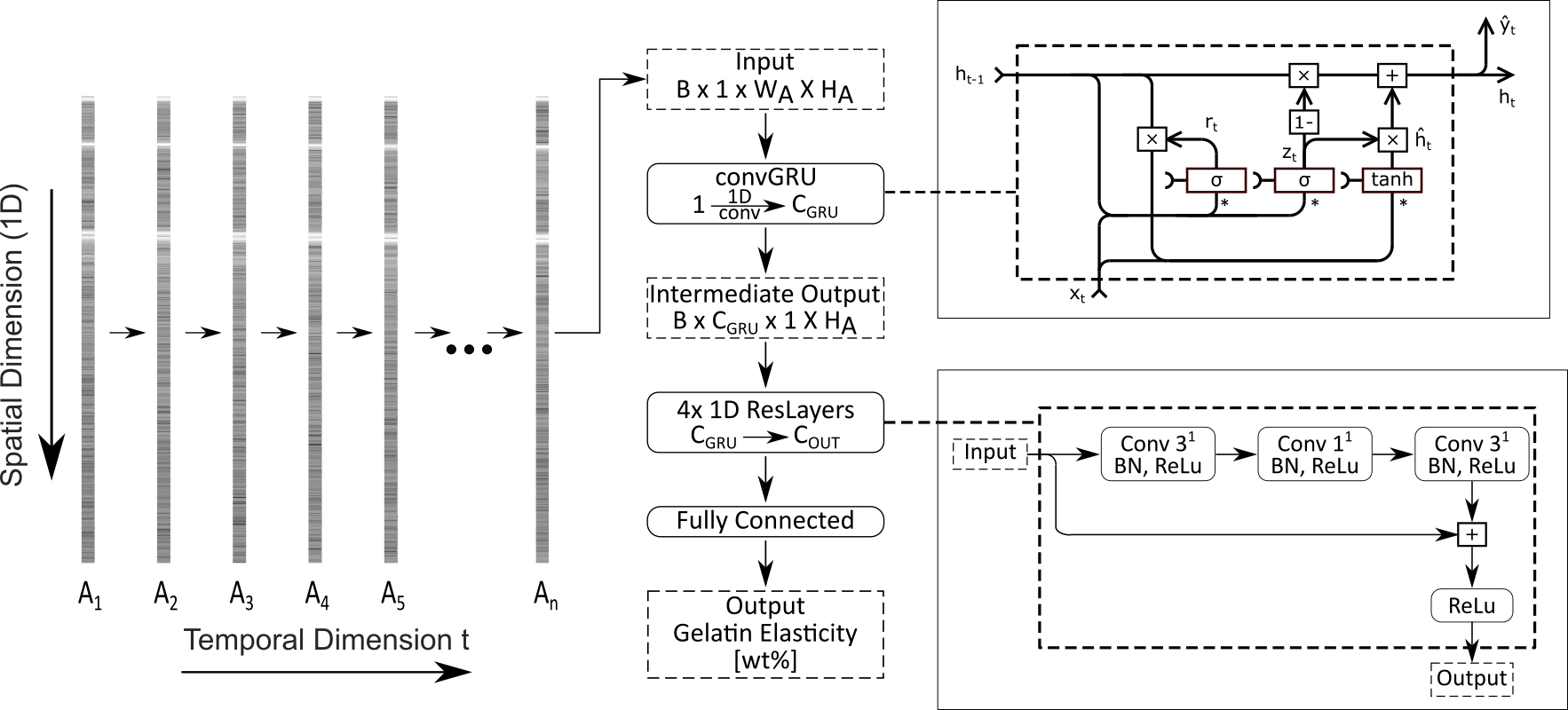}
	\caption{The employed model architecture of the convGRU-CNN sequentially processes each A-scan iteratively to obtain a feature vector. Subsequent spatial processing is conducted in four layers of two residual blocks and the sample is characterized via the gelatin weight ratio.}
	\label{fig:convGRU}
\end{figure*}

\subsection{Network Architectures} 
We compared two network architectures for the mapping of $M_i$ to the gelatin weight ratio. Firstly, we considered the ResNet18-architecture~\cite{resnet2016deep} as our baseline model. And secondly, we employed a convGRU-CNN~\cite{gessert2019spatio} architecture (see Figure~\ref{fig:convGRU}) tailored to the spatio-temporal properties of the OCT input data. A convolutional gated recurrent unit (convGRU) processed the sequential A-scans iteratively and produced a feature vector containing the temporal information of $M_i$. We replaced the dot products in the GRU cell with 1D convolutions such that 
\begin{align*}
    z_t &= \sigma ( W_{hz} * h_{t-1} + W_{xz} * x_{t} + b_z ), \\
    r_t &= \sigma ( W_{hr} * h_{t-1} + W_{xr} * x_{t} + b_r ), \\
    \hat{h}_t &= tanh( W_{h} * (r_t \odot h_{t-1}) + W_{x} * x_{t} + b) \text{ and} \\
    h_t &= (1 - z_t) \odot h_{t-1} + z_t \odot \hat{h}_t
\end{align*}
followed for the update gate $z_t$, the reset gate $r_t$, the candidate activation vector $\hat{h_t}$ and the hidden state $h_t$, respectively. $W$ denotes the trainable filters. The obtained feature vector from the last update of the hidden state $h_n$ was subsequently processed with a 1D CNN architecture to extract the spatial information. It was based on the basic residual building blocks~\cite{resnet2016deep} with 1D convolutions.

\subsection{Data Acquisition} 
Three phantoms were produced for six different elasticities (10wt\%, 12wt\%, 14wt\%, 16wt\%, 18wt\%, 20wt\%). Every phantom of each weight ratio was indented 15 times. The confined compression loading was conducted with randomly chosen loading rates for each indentation experiment between \SI{0.1}{\milli\metre\per\second} and \SI{0.5}{\milli\metre\per\second}. The axial force measurements were employed to maximize compression while preventing surface rupture of the sample. OCT data was acquired during the loading cycle and the OCT intensity data was reconstructed. The indentation data was separated into sequences of $64$ consecutively acquired A-scans representing a sampling frequency of~\SI{86}{\hertz}. The spatio-temporal images were labeled with the corresponding gelatin weight ratio. The obtained data set consisted of approximately \num[tight-spacing = true]{3e6} labeled images $M_i$. We split the data based on the phantoms in individual subsets for training, validation and testing. The validation set was used for the optimization of hyperparameters. The model performance was evaluated based on the mean absolute error (MAE) in wt\% with standard deviation and the correlation coefficient between predictions and labels. We further report the mean inference time in \SI{}{\milli\second} averaged over 300 forward passes. 



%
\begin{table}[b]
	\centering
	\begin{tabular}{c c c c}
		 Model & MAE [wt\%] & CC & IT [\SI{}{\milli\second}]\\ \hline 
		 ResNet18       & $2.16 \pm 1.32$ & 0.8905 &  $\mathbf{3.28 \pm 0.09}$  \\
		 convGRU-CNN    & $\mathbf{1.21 \pm 0.91}$ & \textbf{0.9237} & $34.44 \pm 0.47$ \\
	\end{tabular}
	\caption{MAE, inference time (IT) and correlation coefficient (CC) of the sample characterization for the considered model architectures.}
	\label{tab:res}
\end{table}
\begin{figure}[htb]
    \centering
    \includegraphics[width =0.8\columnwidth]{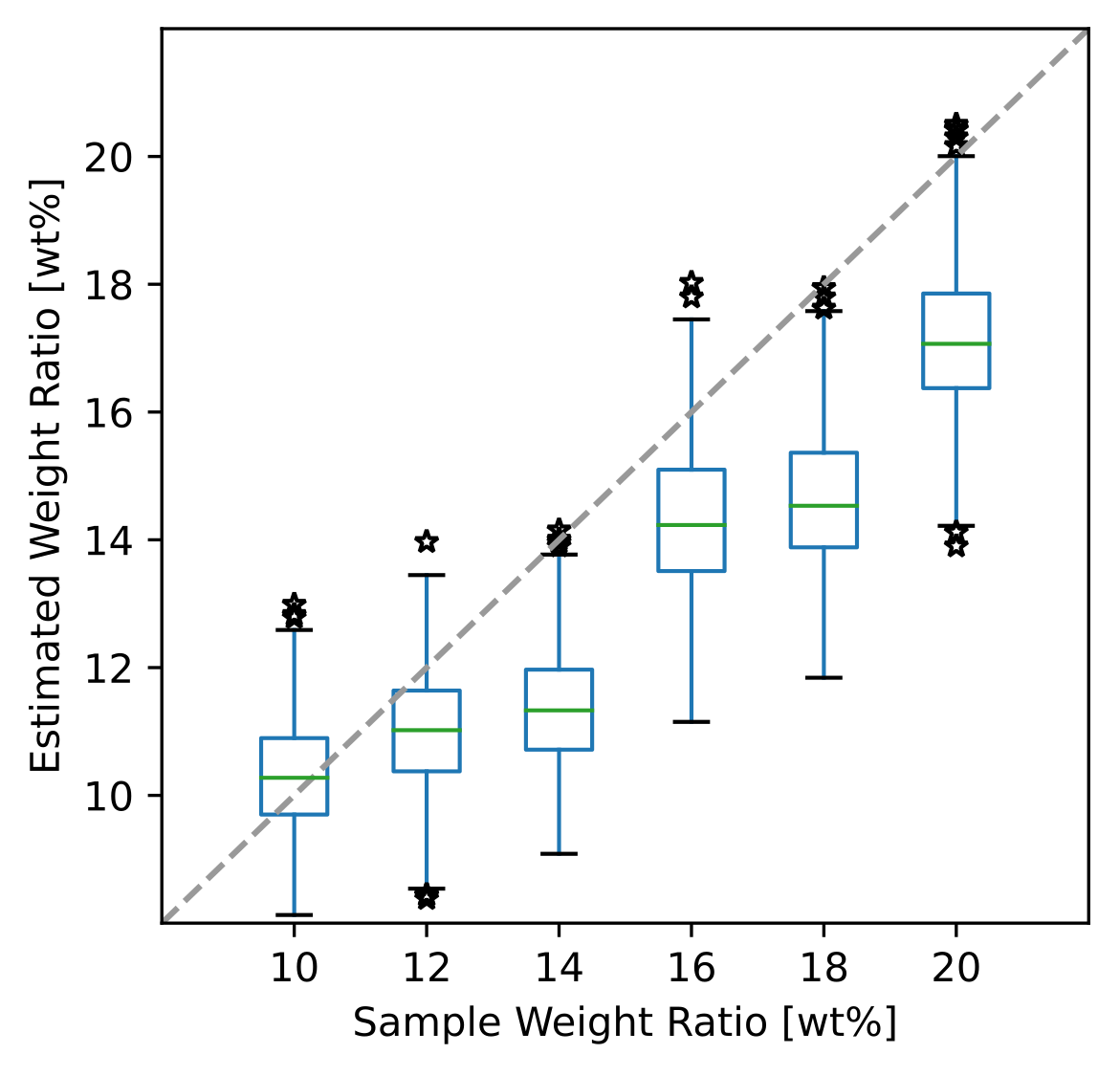}
    \includegraphics[width =0.8\columnwidth]{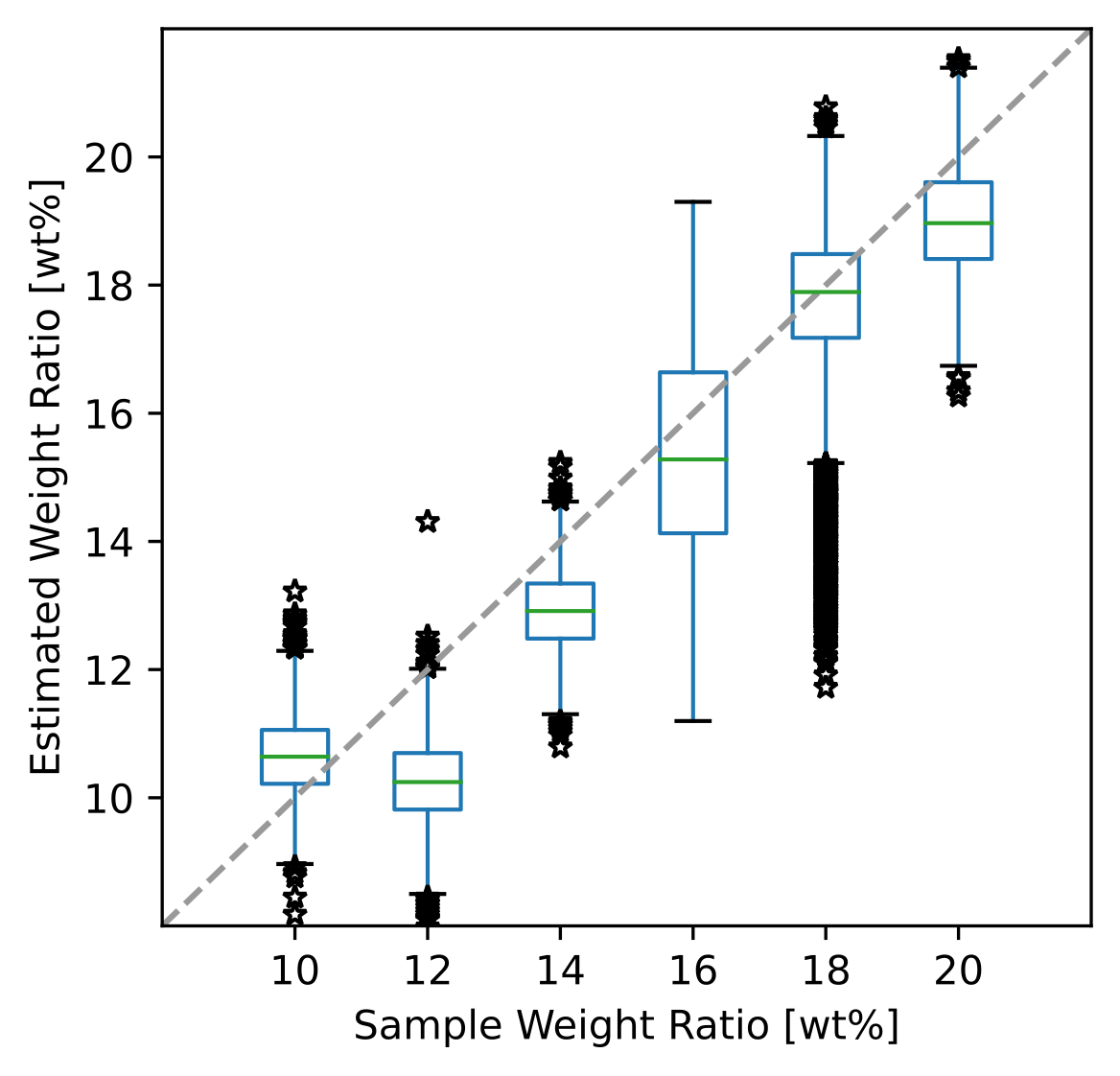}
	\caption{Prediction for the baseline model (top) and convGRU-CNN architecture (bottom) plotted separately over the actual sample elasticity for each input image in the test set.}
	\label{fig:boxplot}
\end{figure}
\section{Results and Discussion}
The results for both model architectures are listed in Table~\ref{tab:res}. The proposed convGRU-CNN and the Resnet18 model resulted in a MAE of $1.21 \pm 0.91$ wt\% and $2.16 \pm 1.32$ wt\%, respectively. The convGRU-CNN also outperformed the baseline model considering the correlation coefficient of $0.9237$ and $0.8905$, respectively. Both models could successfully characterize different sample elasticities in unseen recordings. The indentation loading rate was varied to prevent the models from differentiating the samples based on the change in the epoxy deformation alone. Instead, the models were forced to evaluate the ratio between sample and sensor deformation over the constant temporal sequence of A-scans. The evaluated models were able to characterize the samples and consequently detect the varying responses to the applied mechanical load. The prediction for each input image was also plotted separately for the different sample elasticities and both model types in Figure~\ref{fig:boxplot}. The results show that the baseline model underestimated the elasticity for higher gelatin concentrations. Large deviations of the median from the reference line could be observed and the larger standard deviation of the error indicated a wider spread of the predictions for each weight ratio. The predictions of the convGRU-CNN model were more consistent over the evaluated range and only the elasticity of the 12wt\% gelatin samples was notably underestimated. 
For both models, the Kruskal-Wallis test ($p < 0.0001$) and the afterwards conducted Conover's post hoc test ($p < 0.0001$ for all comparisons) showed that there was a significant difference between the model predictions for each gelatin concentration. 
Regarding the IT, the baseline model provided the faster forward pass while the recurrent nature of the convGRU significantly increased the inference time for the convGRU-CNN model. The lower inference time of the baseline model is therefore of interest for a real-time sample characterization. However, the values are highly hardware (GTX 1080Ti, NVIDIA Corporation, USA) and software (PyTorch) dependent. Alternatively, the more accurate convGRU-CNN model could be utilized with a reduced sampling frequency.

\section{Conclusion}
We proposed a novel OCE needle probe that enabled simultaneous load sensing and OCT imaging. We have demonstrated the application of the needle in indentation experiments on tissue mimicking phantoms. We have shown the end-to-end sample characterization via deep learning directly from the acquired OCT sequences. The samples could be differentiated based on their elastic properties without the need for intermediate calculations of local displacement fields and subsequent strain estimations. The insertion of the needle into deeper sample layers may be realized with an additional guide needle where the probe could enable tissue characterization. This work consequently serves as the foundation towards a deep learning-based quantitative compression-based OCE during biopsies or MIS.


\vspace{\baselineskip}
\noindent \textbf{Author Statement}\\
\textit{Research funding:} This work was partially funded by the TUHH i$^{3}$ initiative and partially by DFG SCHL 1844/2-2.\\
\textit{Conflict of interest:} Authors state no conflict of interest.

\bibliographystyle{abbrvnat}
\bibliography{references}

\begin{thebibliography}{11}
\providecommand{\natexlab}[1]{#1}
\providecommand{\url}[1]{\texttt{#1}}
\expandafter\ifx\csname urlstyle\endcsname\relax
  \providecommand{\doi}[1]{doi: #1}\else
  \providecommand{\doi}{doi: \begingroup \urlstyle{rm}\Url}\fi

\bibitem[Beekmans et~al.(2017)Beekmans, Lembrechts, Van~den Dobbelsteen, and
  Van~Gerwen]{beekmans2017fiber}
S.~Beekmans, T.~Lembrechts, J.~Van~den Dobbelsteen, and D.~Van~Gerwen.
\newblock Fiber-optic fabry-p{\'e}rot interferometers for axial force sensing
  on the tip of a needle.
\newblock \emph{Sensors}, 17\penalty0 (1):\penalty0 38, 2017.

\bibitem[Carrasco-Zevallos et~al.(2017)Carrasco-Zevallos, Viehland, Keller,
  Draelos, Kuo, Toth, and Izatt]{carrasco2017review}
O.~M. Carrasco-Zevallos, C.~Viehland, B.~Keller, M.~Draelos, A.~N. Kuo, C.~A.
  Toth, and J.~A. Izatt.
\newblock Review of intraoperative optical coherence tomography: technology and
  applications.
\newblock \emph{Biomedical optics express}, 8\penalty0 (3):\penalty0
  1607--1637, 2017.

\bibitem[Cui et~al.(2015)Cui, Chang, Kan, Chiorean, Ignee, and
  Dietrich]{cui2015endoscopic}
X.-W. Cui, J.-M. Chang, Q.-C. Kan, L.~Chiorean, A.~Ignee, and C.~F. Dietrich.
\newblock Endoscopic ultrasound elastography: Current status and future
  perspectives.
\newblock \emph{World journal of gastroenterology}, 21\penalty0 (47):\penalty0
  13212, 2015.

\bibitem[Gessert et~al.(2019)Gessert, Priegnitz, Saathoff, Antoni, Meyer,
  Hamann, J{\"u}nemann, Otte, and Schlaefer]{gessert2019spatio}
N.~Gessert, T.~Priegnitz, T.~Saathoff, S.-T. Antoni, D.~Meyer, M.~F. Hamann,
  K.-P. J{\"u}nemann, C.~Otte, and A.~Schlaefer.
\newblock Spatio-temporal deep learning models for tip force estimation during
  needle insertion.
\newblock \emph{International journal of computer assisted radiology and
  surgery}, 14\penalty0 (9):\penalty0 1485--1493, 2019.

\bibitem[Good et~al.(2014)Good, Stewart, Hammer, Scanlan, Shu, Phipps, Reuben,
  and McNeill]{good2014elasticity}
D.~W. Good, G.~D. Stewart, S.~Hammer, P.~Scanlan, W.~Shu, S.~Phipps, R.~Reuben,
  and A.~S. McNeill.
\newblock Elasticity as a biomarker for prostate cancer: a systematic review.
\newblock \emph{BJU international}, 113\penalty0 (4):\penalty0 523--534, 2014.

\bibitem[He et~al.(2016)He, Zhang, Ren, and Sun]{resnet2016deep}
K.~He, X.~Zhang, S.~Ren, and J.~Sun.
\newblock Deep residual learning for image recognition.
\newblock In \emph{Proceedings of the IEEE conference on computer vision and
  pattern recognition}, pages 770--778, 2016.

\bibitem[Kennedy et~al.(2013{\natexlab{a}})Kennedy, Kennedy, and
  Sampson]{kennedy2013review}
B.~F. Kennedy, K.~M. Kennedy, and D.~D. Sampson.
\newblock A review of optical coherence elastography: fundamentals, techniques
  and prospects.
\newblock \emph{IEEE Journal of Selected Topics in Quantum Electronics},
  20\penalty0 (2):\penalty0 272--288, 2013{\natexlab{a}}.

\bibitem[Kennedy et~al.(2013{\natexlab{b}})Kennedy, McLaughlin, Kennedy, Tien,
  Latham, Saunders, and Sampson]{kennedy2013needle}
K.~M. Kennedy, R.~A. McLaughlin, B.~F. Kennedy, A.~Tien, B.~Latham, C.~M.
  Saunders, and D.~D. Sampson.
\newblock Needle optical coherence elastography for the measurement of
  microscale mechanical contrast deep within human breast tissues.
\newblock \emph{Journal of biomedical optics}, 18\penalty0 (12):\penalty0
  121510, 2013{\natexlab{b}}.

\bibitem[Latus et~al.(2017)Latus, Otte, Schl{\"u}ter, Rehra, Bizon,
  Schulz-Hildebrandt, Saathoff, H{\"u}ttmann, and Schlaefer]{latus2017approach}
S.~Latus, C.~Otte, M.~Schl{\"u}ter, J.~Rehra, K.~Bizon, H.~Schulz-Hildebrandt,
  T.~Saathoff, G.~H{\"u}ttmann, and A.~Schlaefer.
\newblock An approach for needle based optical coherence elastography
  measurements.
\newblock In \emph{International Conference on Medical Image Computing and
  Computer-Assisted Intervention}, pages 655--663. Springer, 2017.

\bibitem[Neidhardt et~al.(2020)Neidhardt, Bengs, Latus, Schl{\"u}ter, Saathoff,
  and Schlaefer]{neidhardt20204d}
M.~Neidhardt, M.~Bengs, S.~Latus, M.~Schl{\"u}ter, T.~Saathoff, and
  A.~Schlaefer.
\newblock 4d deep learning for real-time volumetric optical coherence
  elastography.
\newblock \emph{International journal of computer assisted radiology and
  surgery}, pages 1--5, 2020.

\bibitem[Papazoglou et~al.(2012)Papazoglou, Hirsch, Braun, and
  Sack]{papazoglou2012multifrequency}
S.~Papazoglou, S.~Hirsch, J.~Braun, and I.~Sack.
\newblock Multifrequency inversion in magnetic resonance elastography.
\newblock \emph{Physics in Medicine \& Biology}, 57\penalty0 (8):\penalty0
  2329, 2012.

\end{thebibliography}

\end{document}